\documentclass[preprin,showpacs,preprintnumbers,eqsecnum,amsmath,amssymb,nofootinbib]{revtex4}
\usepackage{graphics,epsfig}
\usepackage{graphicx}
\usepackage{dcolumn}
\usepackage{bm}

\begin{document}
\baselineskip=0.8 cm
\title{Slowly rotating Einstein-bumblebee black hole solution and its greybody factor  in a Lorentz violation model}

\author{Chikun Ding$^{1,2,3}$}\thanks{Corresponding author}\email{dingchikun@163.com;  Chikun_Ding@huhst.edu.cn}
\author{Xiongwen Chen$^{1,3}$}\thanks{Corresponding author}\email{chenxiongwen@hhtc.edu.cn}
\affiliation{$^1$Department of Physics, Huaihua University, Huaihua, 418008, P. R. China\\
$^2$Department of Physics, Hunan University of Humanities, Science and Technology, Loudi, Hunan
417000, P. R. China\\
$^3$Key Laboratory of Low Dimensional
Quantum Structures and Quantum Control of Ministry of Education,
and Synergetic Innovation Center for Quantum Effects and Applications,
Hunan Normal University, Changsha, Hunan 410081, P. R. China}

\vspace*{0.2cm}
\begin{abstract}
\baselineskip=0.6 cm
\begin{center}
{\bf Abstract}
\end{center}

We obtain an exact slowly rotating Einstein-bumblebee black hole solution by solving the corresponding $rr$ and $t\phi$ components of the gravitational field equations in both cases: A, $b_\mu=(0,b(r),0,0)$; B, $b_\mu=(0,b(r),\mathfrak{b}(\theta),0)$. Then we check the other  gravitational field equations and the bumblebee field motion equations by using this solution. We find that in the case A, there exists a slowly rotating black hole solution indeed for arbitrary LV (Lorentz violation) coupling constant $\ell$; however as in the case B, there exists this slowly rotating solution if and only if the coupling constant $\ell$ is as small as or smaller than the angular momentum $a$. Till now there seems to be no full rotating black hole solution, so one can't use the Newman-Janis algorithm to generate a rotating solution in Einstein-bumblebee theory. It is similar as that in Einstein-aether theory where there exists only some slowly rotating black hole solutions.
In order to study the effects of this Lorentz symmetry broken, we consider the black hole greybody factor and find that when angular index $l=0$, the LV constant $\ell$ decreases the effective potential and enhances the absorption probability, which is similar to that of the non-minimal derivative coupling theory.

\end{abstract}

\pacs{ 04.50.Kd, 04.20.Jb, 04.70.Dy  } \maketitle

\vspace*{0.2cm}
\section{Introduction}

Lorentz invariance(LI) is a most fundamental principle of general relativity (GR) and the standard model(SM) of particle physics. It is therefore no surprise that most theories of gravity encompass this symmetry and little attention has been paid in understand the implications of the breaking of LI.  However, LI should not be an exact principle at all energy scales  \cite{mattingly}, for example, when one considers the unity of quantum mechanics and GR, it should not be applicable. Both GR and SM based on LI and the background of spacetime, but they process their problems in profoundly different ways. GR is a classical field theory in curved spacetime that ignores all quantum features of particles; SM is a quantum field theory in flat spacetime that forgets all gravitational effects of particles. For collisions of particles of $10^{30}$ eV energy (energy higher than Planck scale), the gravitational interactions estimated by GR are very powerful and gravity should not be ignorant\cite{camelia}. So in this very high energy scale,  one have to reconsider combining SM with GR in an unitary theory, etc., ``quantum gravity".  Therefore, the study of Lorentz violation (LV) is an useful approach to investigate the essences of modern physics. These searches invole LV in the neutrino sector \cite{dai}, the standard-model extension (SME) \cite{colladay}, LV in the non-gravity sector \cite{coleman}, and LV effect on the creation of atmospheric showers \cite{rubtsov}.

Experimental confirmation of this idea of quantum gravity is challenging because direct experiments at the Plank scale are impractical. However, suppressed effects emerging from the underlying unified quantum gravity theory might be observable at our low energy scale. So that the search for reminiscent quantum gravity effects at low energy regime has attracted attention over the last decades. The combination of GR and SM provides a remarkably successful description of nature.
The SME is an effective field theory which studies gravity and the SM at low energy scales. And it picturing the SM coupled to GR, enabling dynamical curvature modes, and involves extra items embracing information about the LV happening at the Plank scale \cite{kostelecky2004}. The LV items in the SME have the form of Lorentz-violating operators coupled to coefficients with Lorentz indices. The appearance of LV in a local Lorentz frame is shown by a nonzero vacuum value for one or more quantities taking along local Lorentz indices. A specific theory is the ``bumblebee" pattern, where the LV arouses from the dynamics of a single vector or axial-vector field $B_\mu$, known as the bumblebee field. This model is a simple effective theory of gravity with LV in the SME and a subset of Einstein-aether theory\cite{guiomar,ding2015,ding2016}. It is controlled by a potential revealing a minimum scrolls to its vacuum expectation value showing that the vacuum of the theory obtains a preferential direction in the spacetime. Bumblebee gravitational model was first studied by Kostelecky and Samuel in 1989 \cite{dickinson,kostelecky1989} as a specific pattern for unprompted Lorentz violation.

Deriving black hole solutions are very requisite  tasks in any theory of gravity, due to that they give a large amount of information about the quantum gravity area. In 2018, R. Casana {\it et al} gave an exact Schwarzschild-like solution in this bumblebee gravity model and considered its some classical tests \cite{casana}. Then Rong-Jia Yang {\it el al} researched the accretion onto this black hole \cite{yang} and discovered that the LV parameter $\ell$ will decrease the mass
accretion rate. The rotating black hole solutions are the most relational subsets for astrophysics. In 2020, C. Ding {\it et al} found an exact Kerr-like solution through solving Einstein-bumblebee gravitational field equations and studied its black hole shadow\cite{ding2020}. However, this solution doesn't seem to meet the bumblebee field motion equation. So in the present paper, we try to seek a slowly rotating black hole solution in both cases: $b_\mu=(0,b(r),0,0)$ and $b_\mu=(0,b(r),\mathfrak{b}(\theta),0)$.

We then study black hole greybody factor and obtain some deviations from GR and some LV gravity theories. The rest of the paper is arranged as following. In Sec. II we give the background for the Einstein-bumblebee theory. In Sec. III, we give the slowly rotating black hole solution by solving the gravitational field equations. In Sec. IV, we study its black hole greybody factor and find some effects of the Lorentz breaking constant $\ell$. Sec. V is for a summary.

\section{Einstein-bumblebee theory}

In the bumblebee gravity theory, the bumblebee vector field $B_{\mu}$ gets a nonzero vacuum expectation value, via a given potential, leading a spontaneous Lorentz symmetry breaking in the gravitational sector. Its
 action\cite{bluhm} is,
 \begin{eqnarray}
&&\mathcal{S}=
\int d^4x\sqrt{-g}\Big[\frac{1}{16\pi G_N}(\mathcal{R}-2\Lambda+\varrho B^{\mu}B^{\nu}\mathcal{R}_{\mu\nu}+\sigma B_\mu B^\mu \mathcal{R})-\frac{\tau_1}{4}B^{\mu\nu}B_{\mu\nu}+\frac{\tau_2}{2}D_\mu B_\nu D^\mu B^\nu
\nonumber\\
&&\quad\quad +\frac{\tau_3}{2}D_\mu B^\mu D_\nu B^\nu-V(B_\mu B^{\mu}\mp b^2)+\mathcal{L}_M\Big], \label{action0}
\end{eqnarray}
in which $b^2$ is a real positive constant. Lorentz and/or $CPT$ (charge conjugation, parity and time reversal) violation is opened by the potential $V(B_\mu B^{\mu}\mp b^2)$. It gives a nonzero vacuum expectation value (VEV) for bumblebee field $B_{\mu}$ implying that the vacuum of this theory gets a preferential direction in the spacetime. This potential is assumed to have a minimum at $B^{\mu}B_{\mu}\pm b^2=0$ and $V'(b_{\mu}b^{\mu})=0$ to assure the breaking of the $U(1)$ symmetry, where the field $B_{\mu}$ obtains a nonzero VEV, $\langle B^{\mu}\rangle= b^{\mu}$. The vector $b^{\mu}$ is a function of the spacetime coordinates and has constant value $b_{\mu}b^{\mu}=\mp b^2$, where $\pm$ signs imply that $b^{\mu}$ is timelike or spacelike, respectively.
The bumblebee field strength is
\begin{eqnarray}
B_{\mu\nu}=\partial_{\mu}B_{\nu}-\partial_{\nu}B_{\mu}.
\end{eqnarray}
The real constants $\varrho,\;\sigma,\;\tau_1,\;\tau_2,\;\tau_3 $ determine the form of the kinetic terms for the bumblebee field. The term $\mathcal{L}_M$ represents possible interactions with matter or external currents. It should be note that if $\varrho=\sigma=0$ and with linear Lagrange-multiplier potential
\begin{eqnarray}
V=\lambda (B_\mu B^{\mu}\mp b^2),
\end{eqnarray}
this bumblebee model becomes the special case of Einstein-aether theory\cite{bluhm}. In Einstein-aether theory \cite{ding2015,ding2016}, the Lorentz symmetry is broken by an introduced  tensor field $u^a$ with the constraint $u_au^a=-1$, termed aether, which is timelike everywhere and everytime. Then there exists  a preferred time direction at every point of spacetime, i.e.,  a preferred frame of reference. The introduction of the aether vector allows for some novel effects, e.g., matter fields can travel faster than the speed of light, dubbed superluminal particle. In Ref. \cite{ding2015}, we obtained a series of charged Einstein-aether black hole solutions in 4 dimensional spacetime and studied their Smarr formula; In Ref. \cite{ding2016}, we obtained a series of neutral and charged black hole solutions in 3 dimensional spacetime.

In this study, the constant $\tau_1=1$, $\Lambda=\sigma=\tau_2=\tau_3=0$ and no $\mathcal{L}_M$, i.e.,
\begin{eqnarray}
\mathcal{S}=
\int d^4x\sqrt{-g}\Big[\frac{1}{16\pi G_N}(\mathcal{R}+\varrho B^{\mu}B^{\nu}\mathcal{R}_{\mu\nu})-\frac{1}{4}B^{\mu\nu}B_{\mu\nu}
-V(B_\mu B^{\mu}\mp b^2)\Big], \label{action}
\end{eqnarray}
where $\varrho$ dominates the non-minimal gravity interaction to bumblebee field $B_\mu$.
The action (\ref{action}) yields the gravitational field equation in vacuum
\begin{eqnarray}\label{einstein0}
\mathcal{R}_{\mu\nu}-\frac{1}{2}g_{\mu\nu}\mathcal{R}=\kappa T_{\mu\nu}^B,
\end{eqnarray}
where $\kappa=8\pi G_N$ and the bumblebee energy momentum tensor $T_{\mu\nu}^B$ is
\begin{eqnarray}\label{momentum}
&&T_{\mu\nu}^B=B_{\mu\alpha}B^{\alpha}_{\;\nu}-\frac{1}{4}g_{\mu\nu} B^{\alpha\beta}B_{\alpha\beta}- g_{\mu\nu}V+
2B_{\mu}B_{\nu}V'\nonumber\\
&&+\frac{\varrho}{\kappa}\Big[\frac{1}{2}g_{\mu\nu}B^{\alpha}B^{\beta}R_{\alpha\beta}
-B_{\mu}B^{\alpha}R_{\alpha\nu}-B_{\nu}B^{\alpha}R_{\alpha\mu}\nonumber\\
&&+\frac{1}{2}\nabla_{\alpha}\nabla_{\mu}(B^{\alpha}B_{\nu})
+\frac{1}{2}\nabla_{\alpha}\nabla_{\nu}(B^{\alpha}B_{\mu})
-\frac{1}{2}\nabla^2(B^{\mu}B_{\nu})-\frac{1}{2}
g_{\mu\nu}\nabla_{\alpha}\nabla_{\beta}(B^{\alpha}B^{\beta})\Big].
\end{eqnarray}
The prime denotes differentiation with respect to the argument,
\begin{eqnarray}
V'=\frac{\partial V(x)}{\partial x}\Big|_{x=B^{\mu}B_{\mu}\pm b^2}.
\end{eqnarray}
Using the trace of Eq. (\ref{einstein0}), we obtain the trace-reversed version
\begin{eqnarray}\label{einstein}
\mathcal{R}_{\mu\nu}=\kappa T_{\mu\nu}^B+2\kappa g_{\mu\nu}V
-\kappa g_{\mu\nu} B^{\alpha}B_{\alpha}V'+\frac{\varrho}{4}g_{\mu\nu}\nabla^2(B^{\alpha}B_{\alpha})
+\frac{\varrho}{2}g_{\mu\nu}\nabla_{\alpha}\nabla_{\beta}(B^{\alpha}B^{\beta}).
\end{eqnarray}

The equation of motion for the bumblebee field is
\begin{eqnarray}\label{motion}
\nabla ^{\mu}B_{\mu\nu}=2V'B_\nu-\frac{\varrho}{\kappa}B^{\mu}R_{\mu\nu}.
\end{eqnarray}

In the following, we suppose that the bumblebee field is frosted at its VEV, i.e., it is
\begin{eqnarray}
B_\mu=b_\mu,
\end{eqnarray}
then the specific form of the potential controlling its dynamics is irrelevant.
And  as a result, we have $V=0,\;V'=0$. Then the first two terms in Eq. (\ref{momentum}) are like those of the electromagnetic field, the only distinctive are the coupling items to Ricci tensor. Under this condition,  Eq. (\ref{einstein}) leads to gravitational field equations
\begin{eqnarray}\label{bar}
\bar R_{\mu\nu}=0,
\end{eqnarray}
with
\begin{eqnarray}\label{barb}
&&\bar R_{\mu\nu}=\mathcal{R}_{\mu\nu}-\kappa b_{\mu\alpha}b^{\alpha}_{\;\nu}+\frac{\kappa}{4}g_{\mu\nu} b^{\alpha\beta}b_{\alpha\beta}+\varrho b_{\mu}b^{\alpha}\mathcal{R}_{\alpha\nu}
+\varrho b_{\nu}b^{\alpha}\mathcal{R}_{\alpha\mu}
-\frac{\varrho}{2}g_{\mu\nu}b^{\alpha}b^{\beta}\mathcal{R}_{\alpha\beta}+\bar B_{\mu\nu},\nonumber\\
&&\bar B_{\mu\nu}=-\frac{\varrho}{2}\Big[
\nabla_{\alpha}\nabla_{\mu}(b^{\alpha}b_{\nu})
+\nabla_{\alpha}\nabla_{\nu}(b^{\alpha}b_{\mu})
-\nabla^2(b_{\mu}b_{\nu})\Big].
\end{eqnarray}
In the next section, we derive the slowly rotating black hole solution by solving  gravitational equations in this Einstein-bumblebee model.

\section{Slowly rotating solution in Einstein-bumblebee model}
In this section, we will find the slowly rotating black hole solution through solving Einstein-bumblebee gravitational equations.
Rotating black hole solutions are the most important for astrophysics.  However, the deriving an exact rotating solution is very troublesome. For example, Schwarzschild black hole solution was obtained in 1916 soon after GR was announced \cite{sch}. However, till 47 years later, in 1963, the rotating counterpart was appeared\cite{kerr}.
So many scholars use the Newman-Janis algorithm \cite{newman} to obtain a full\footnote{The term ``full" means that there is no man-made restrictions on the angular momentum $a$.} rotating black hole solution and haven't check  the gravitational field equations with this obtained solution. Some authors have proved this method can not work with nonlinear source \cite{lammerzahl}.

In Ref. \cite{ding2020}, we have found the exact Kerr-like solution in the case that bumblebee field $b_\mu=(0,b(r,\theta),0,0)$ by solving gravitational  field equations. However, this full rotating solution does't seem to meet the bumblebee field equation.
And when we consider the case that $b_\mu=(0,b(r),\mathfrak{b}(\theta),0)$, it is very difficult and seems impractical by the same method. So we here find the slowly rotating black hole solution.
The slowly rotating stationary axially symmetric black hole metric  have the general form
\begin{eqnarray}\label{metric}
&&ds^2=-U(r)dt^2+\frac{1+\ell}{U(r)}dr^2+2F(r)H(\theta)adtd\phi+r^2d\theta^2
+r^2\sin^2\theta d\phi^2+\mathcal{O}(a^2),
\end{eqnarray}
where $a$ is a small constant denoting the rotating angular momentum, and $\mathcal{O}(a^2)$ denotes a little quantity as small as or smaller than the second order of $a$, which can be ignored here. We will use this metric ansatz to set up gravitational field equations.

In this study, we focus on that the bumblebee field acquiring a radial vacuum energy expectation since the spacetime curvature has a strong radial variation when compared with very slow temporal changes. So the bumblebee field is spacelike($b_\mu b^\mu=$ positive constant) and assumed to be
\begin{eqnarray}
\text{Case A:}\;b_\mu=\big(0,b(r),0,0\big);\quad\text{Case B:}\;b_\mu=\big(0,b(r),\mathfrak{b}(\theta),0\big).
\end{eqnarray}
The case A is considered by Casana and Ding {\it et al} in Ref. \cite{casana,ding2020} for bumblebee field coupling to the gravitational field, and the case B is considered by Chen {\it et al} in Ref. \cite{chen2020} for bumblebee field coupling to the electromagnetic field.
Then the bumblebee field strength is
\begin{eqnarray}
b_{\mu\nu}=\partial_{\mu}b_{\nu}-\partial_{\nu}b_{\mu},
\end{eqnarray}
whose components are all zero for the case A and B. And their divergences are all zero, i.e.,
\begin{eqnarray}
\nabla^{\mu}b_{\mu\nu}=0.
\end{eqnarray}
From the equation of motion (\ref{motion}), we have
\begin{eqnarray}
b^{\mu}\mathcal{R}_{\mu\nu}=0\label{motion2}.
\end{eqnarray}
The gravitational field equations (\ref{bar}) become
\begin{eqnarray}\label{}
&&\mathcal{R}_{\mu\nu}+\bar B_{\mu\nu}=0.
\end{eqnarray}
The explicit form of $b_\mu$ is
\begin{eqnarray}\label{bu}
\text{Case A:}\;b_\mu=\Big(0,b_0\sqrt{\frac{1+\ell}{U(r)}},0,0\Big);\quad\text{Case B:}\;b_\mu=\Big(0,b_0\sqrt{\frac{1+\ell}{U(r)}}\;,ab_0\cos\theta,0\Big),
\end{eqnarray}
where $b_0$ is a real constant. The amplitude of these bumblebee field is
\begin{eqnarray}\label{}
b_\mu b^{\mu}=g^{\mu\nu}b_\mu b_{\nu}=b_0^2,
\end{eqnarray}
in the case A; for the case B, it is
\begin{eqnarray}\label{}
b_\mu b^{\mu}=g^{\mu\nu}b_\mu b_{\nu}=b_0^2+\mathcal{O}(a^2),
\end{eqnarray}
which are both consistent with the condition $b_\mu b^\mu=$ positive constant.

 For the metric (\ref{metric}), the nonzero components of Ricci tensor are $\mathcal{R}_{tt},\mathcal{R}_{t\phi},\mathcal{R}_{rr},
 \mathcal{R}_{r\theta},\mathcal{R}_{\theta\theta},\mathcal{R}_{\phi\phi}$, shown in the appendix.
 We consider the following both gravitational field equations (in both cases A and B)
\begin{eqnarray}
&&\mathcal{R}_{rr}+\bar B_{rr}=-\frac{1}{2rU}(rU''+2U')+\mathcal{O}(a^2)=0,\label{rr}\\
&&\mathcal{R}_{t\phi}+\bar B_{t\phi}=-\frac{a}{2}\left(HUF''+2FH\frac{U'}{r}+\frac{F}{r^2}H''
-\frac{F\cos\theta}{r^2\sin\theta}H'\right)+\mathcal{O}(a^2)=0,\label{tp}
\end{eqnarray}
where the prime $'$ is the derivative with respect to the corresponding argument, respectively.
From the Eq. (\ref{rr}), one can obtain the function $U(r)$,
\begin{eqnarray}
U=-\frac{C_1}{r}+C_2,
\end{eqnarray}
where $C_1,\;C_2$ are constants. By using the condition of asymptotically flat, one will chooses $C_2=1$ and $C_1=2M$, where $M$ is the mass of the black hole,
\begin{eqnarray}
U=1-\frac{2M}{r}.
\end{eqnarray}
From the Eq. (\ref{tp}), one can obtain the function $F(r)$ and $H(\theta)$,
\begin{eqnarray}
F=\frac{2M}{r}, \;H=\sin^2\theta.
\end{eqnarray}

Lastly, substituting these quantities into Eqs. (\ref{metric}) and (\ref{bu}), we can get the bumblebee field $b_\mu=(0,b_0\sqrt{(1+\ell)r/(r-2M)},0,0)$ for the case A, $b_\mu=(0,b_0\sqrt{(1+\ell)r/(r-2M)},ab_0\cos\theta,0)$ for the case B, and the
slowly rotating metric in the bumblebee gravity in both cases is
\begin{eqnarray}\label{bmetric}
ds^2=- \Big(1-\frac{2M}{r}\Big)dt^2-\frac{4Ma\sin^2\theta}{r}
dtd\varphi+\frac{(1+\ell)r}{r-2M}dr^2+r^2d\theta^2
+r^2\sin^2\theta d\phi^2.
\end{eqnarray}

If $\ell\rightarrow0$, it recovers the usual slowly rotating Kerr metric.
When $a\rightarrow0$, it becomes
\begin{eqnarray}\label{}
ds^2=- \Big(1-\frac{2M}{r}\Big)dt^2+\frac{1+\ell}{1-2M/r}dr^2+r^2d\theta^2
+r^2\sin^2\theta d\varphi^2,
\end{eqnarray}
which is the same as that in Ref. \cite{casana}.
The metric (\ref{bmetric}) represents a Lorentz-violating  black hole solution with the slowly rotating angular momentum $a$. It is easy to see that horizon locates at $r_+=2M$.

Nextly, we consider the bumblebee motion equation and check for other gravitational equations. From the bumblebee field motion equation (\ref{motion2}), one can obtain the following both equations
\begin{eqnarray}
b^r\mathcal{R}_{rr}+b^\theta\mathcal{R}_{\theta r}=0,\; b^r\mathcal{R}_{r\theta}+b^\theta\mathcal{R}_{\theta \theta}=0.
\end{eqnarray}
From the appendix, we have $\mathcal{R}_{rr}=-(rU''+2U')/2rU+\mathcal{O}(a^2)=0+\mathcal{O}(a^2)$, $\mathcal{R}_{r\theta}=0+\mathcal{O}(a^2)$ and $\mathcal{R}_{\theta \theta}=-(rU'+U)/(1+\ell)+1+\mathcal{O}(a^2)=\ell/(1+\ell)+\mathcal{O}(a^2)$. Then one can see that the first equation is fulfilled for both cases; the second one can be fulfilled in the case A (due to that $b^\theta=0$). However, the second equation in the case B is
\begin{eqnarray}
b^\theta\mathcal{R}_{\theta \theta}=\frac{b_0\cos\theta}{1+\ell}\cdot\frac{a\ell}{r^2}+\mathcal{O}(a^2).
\end{eqnarray} If and only if the coupling constant $\ell$ is also smaller enough as angular momentum $a$, can the second motion equation be fulfilled in the case B. As for the other gravitational equations, in the case A, they are,
\begin{eqnarray}
&&\mathcal{R}_{tt}+\bar B_{tt}=0
+\mathcal{O}(a^2),\\
&&\mathcal{R}_{r\theta}+\bar B_{r\theta}=0+\mathcal{O}(a^2),
\\&&\mathcal{R}_{\theta\theta}+\bar B_{\theta\theta}=0+\mathcal{O}(a^2),
\\&&\mathcal{R}_{\phi\phi}+\bar B_{\phi\phi}=0+\mathcal{O}(a^2),
\end{eqnarray}
which  are all fulfilled.
However in the case B, there have similar limits onto these equations to be fulfilled,
\begin{eqnarray}
&&\mathcal{R}_{tt}+\bar B_{tt}=\frac{\cos2\theta}{2\sqrt{1+\ell}\sin\theta}\cdot\frac{a\ell\sqrt{U}}{r^2}
+\mathcal{O}(a^2),\\
&&\mathcal{R}_{r\theta}+\bar B_{r\theta}=\frac{\cos\theta}{\sqrt{1+\ell}}\cdot\frac{a\ell}{r^2\sqrt{U}}+\mathcal{O}(a^2),
\\&&\mathcal{R}_{\theta\theta}+\bar B_{\theta\theta}=\frac{1}{\sqrt{1+\ell}\sin\theta}\cdot\frac{a\ell}{r\sqrt{U}}
(r\sin^2\theta U'-\cos2\theta U)+\mathcal{O}(a^2),
\\&&\mathcal{R}_{\phi\phi}+\bar B_{\phi\phi}=\frac{\sin\theta}{2\sqrt{1+\ell}}\cdot\frac{a\ell}{r\sqrt{U}}
(r\cos^2\theta U'+2\cos2\theta U)+\mathcal{O}(a^2).
\end{eqnarray}

In conclusion, there exists a slowly rotating black hole solution in the case A for arbitrary coupling constant $\ell$; however for the case B, there exist a slowly rotating black hole solution if and only if the coupling constant is smaller enough as the rotating angular momentum $a$. It is interesting that in both cases, the forms of the solution are the same. Till now, there seems to be no full rotating black hole solution. Because that for the case A, when the slowly rotating restrictions on $a$ is released, it becomes $b_\mu=(0,b(r,\theta),0,0)$, and the obtained full rotating solution \cite{ding2020} doesn't seem to meet the bumblebee field equation; for the case B, it is an open issue that whether there exist a full rotating solution when the coupling constant $\ell$ is small enough. Therefore one can't use Newman-Janis algorithm to obtain a full rotating black hole solution. This is similar to Einstein-aether theory, where there exist only a slowly rotating black hole solution\cite{barausse} with a spherically
symmetric (hypersurface-orthogonal) aether field configuration in the case of $c_{14}=0, c_{123}\neq0$ \cite{tao}.

\section{greybody factor}
The above slowly rotating solution contains the effects of the bumblebee field and can be used
to study the effects of the bumblebee field in the black hole greybody factor.
In this section, we study some observational signatures on the Lorentz-violating parameter $\ell$ by analyzing black hole greybody factor (Hawking radiation) with the metric (\ref{bmetric}), and try to find some deviation from GR and some similarities to other LV black holes.

In Ref. \cite{ding2010p}, we obtained an analytical expressions for the greybody factor and dynamic evolution for the scalar field in the Ho\u{r}ava-Lifshitz black hole. In Ref. \cite{ding2010j}, we studied the greybody factor of the slowly rotating Kerr-Newman black hole in a non-minimal derivative coupling theory. In this model, the kinetic term of the scalar field $\psi$ only coupled with the Einstein's tensor, $G^{\mu\nu}\partial_\mu\psi\partial_\nu\psi$. This coupling is confirmed to be breaking Lorentz symmetry\cite{chen2015}.

The  Klein-Gordon equation in the Einstein-bumblebee black hole spacetime (the scalar field coupling to the bumblebee field is ignored here) is
\begin{eqnarray}
\frac{1}{\sqrt{-g}}\partial_{\mu}\bigg(\sqrt{-g}g^{\mu\nu}\partial_{\nu}\psi\bigg) =0.\label{WE}
\end{eqnarray}
Applying the spherical harmonics
\begin{equation}
\psi(t,r,\theta,\varphi)= e^{-i\omega t}\,e^{i m \varphi}\,R_{\omega lm}(r)
\,T^{m}_{l}(\theta, a \omega)\,,
\end{equation}
and substituting the metric (\ref{bmetric}) into Eq. (\ref{WE}), we can
 get the following radial equation
\begin{equation}
\frac{d}{dr}\biggl[\big(r^2-2Mr\big)\,\frac{d R_{\omega l m}}{dr}\biggr]+
(1+\ell)\left[\frac{r^2\omega(r^2\omega-2am)}{r^2-2Mr}-l(l+1)+2am\omega
\right]R_{\omega l m}=0\,. \label{radial}
\end{equation}
 Before attempting to solve it analytically, we first
analyze the profile of effective potential which characterizes the emission
process. Defining a new radial function
\begin{equation}
R_{\omega l m}(r)=\frac{\tilde{R}_{\omega l m}(r)}{r},
\end{equation}
and useing the tortoise coordinate $x$ as following
\begin{equation}
\frac{d}{dx}=\big(1-\frac{2M}{r}\big)\frac{d}{dr},
\end{equation}
Eq.(\ref{radial}) can be rewritten in the
standard Schr\"{o}dinger equation as
\begin{equation}
\Big(\frac{d^2}{dx^2}-V_{eff}\Big)\tilde{R}_{\omega l m}(x)=0,
\end{equation}
where the effective potential is
\begin{equation}
V_{eff}=(1+\ell)\Big(-\omega^2+\frac{4M}{r^3}ma\omega\Big)+\big(1-\frac{2M}{r}\big)
\Big[\frac{4M}{r^3}+(1+\ell)\frac{l(l+1)}{r^2}\Big].
\end{equation}
 For
graphical analysis, we display the dependence of effective potential on different parameters in  Fig. 1.
\begin{figure}[ht]\label{figf}
\begin{center}
\includegraphics[width=8.0cm]{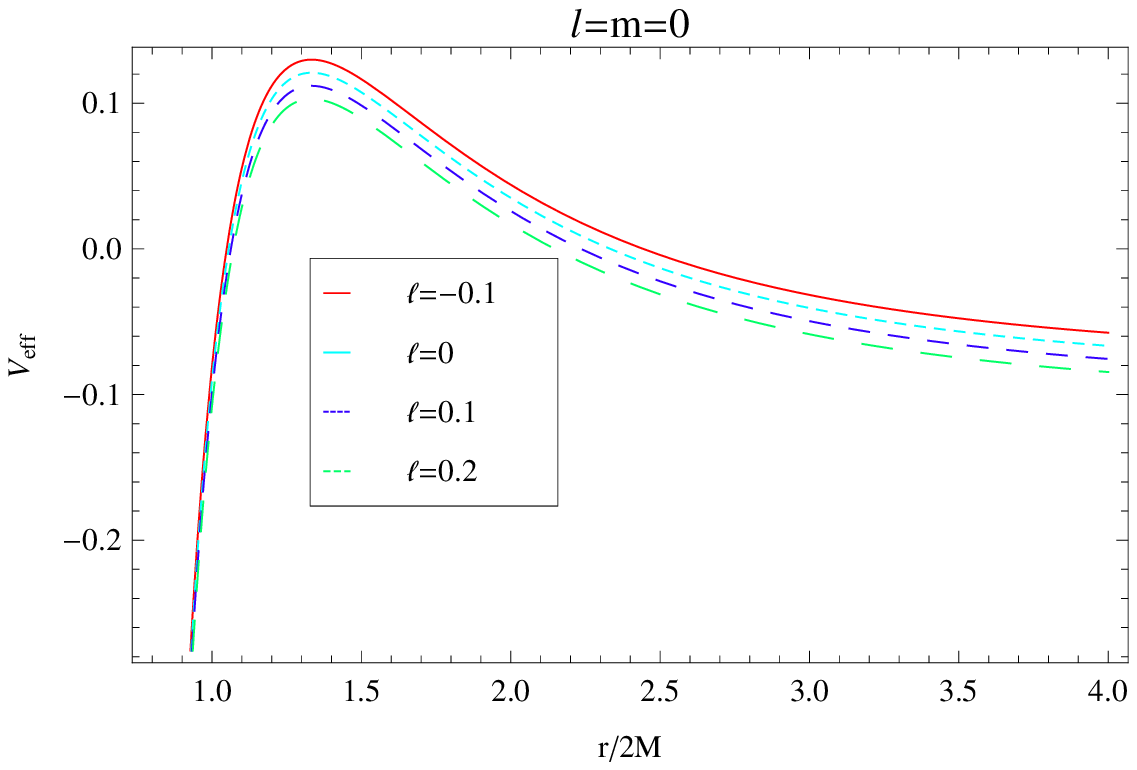}\;\;\;\;\includegraphics[width=8.0cm]{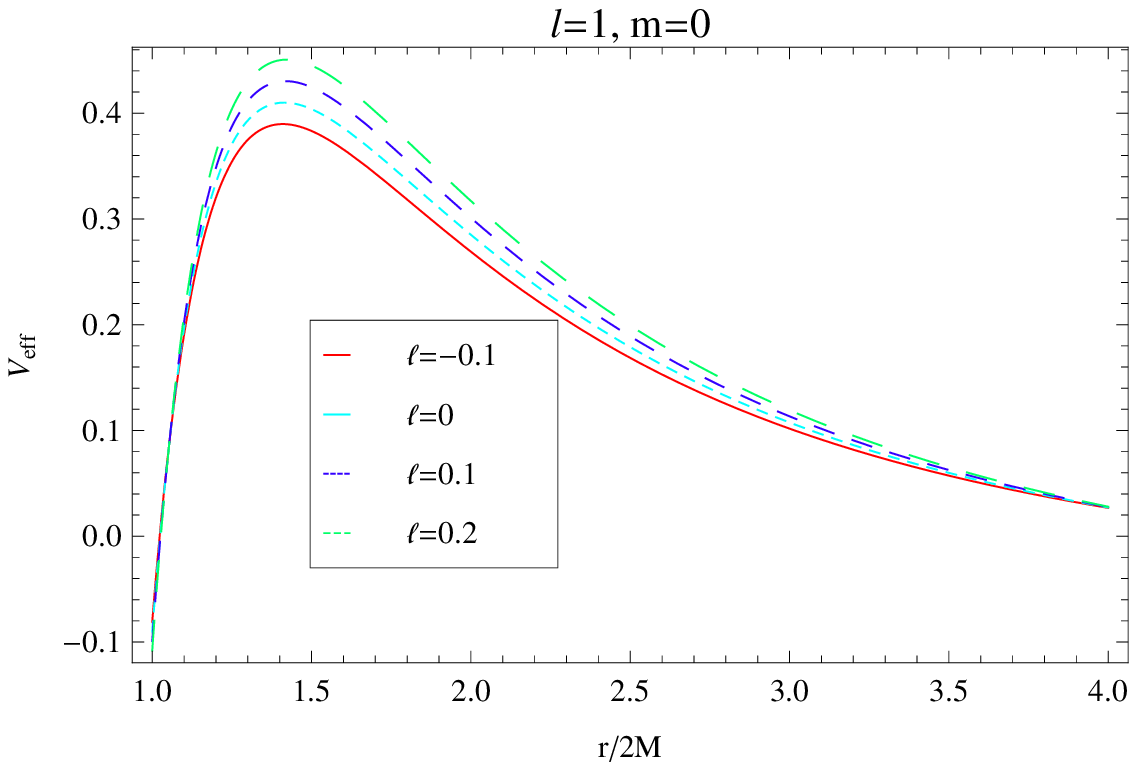}
\caption{Variety of the potential $V_{eff}$ with different coupling constant $\ell$ of a
 scalar field in the slowly rotating Einstein-bumblebee black hole for
 fixed quantity $\omega=0.3$. }
\end{center}
\end{figure}
 It is found that the gravitational
barrier decreases gradually with the increase in the LV coupling constant $\ell$ when $l=0$ (left plot of
Fig. 1), which is similar to that of Einstein-aether theory \cite{tao} and non-minimal coupling theory\cite{ding2010j}. However, the barrier increases with $\ell$ when $l=1$ (right plot of
Fig. 1).  Increasing the effective potential means
reducing the emission of scalar fields, so the LV coupling constant $\ell$ will affect the black hole greybody factor.

Now, we give an analytical solution to Eq. (\ref{radial}). We can change the radial variable $r$
\cite{ding2010p,ding2010j} to the function $f(r)$ as
\begin{eqnarray}
r \rightarrow f(r) = 1-\frac{2M}{r} \,\,\Longrightarrow\,
\frac{d }{dr}=\frac{1-f}{r}\frac{d}{df}\,,
\end{eqnarray}
then the equation (\ref{radial}) can be
transformed into
\begin{eqnarray}
f(1-f)\frac{d^2R(f)}{d f^2}+(1-f)\frac{d R(f)}{d f}
+\bigg[\frac{K^2_*}{(1-f)f}
-\frac{\Lambda^m_l}{(1-f)}\bigg]R(f)=0,\label{r1}
\end{eqnarray}
where \footnote{For the sake of solving the  equation mathematically, we
should make all coefficients in it dimensionless, therefore we define
these quantities.}
\begin{eqnarray}
 K_*=\big(\omega  r_+-a_*m\big)\sqrt{1+\ell},\;a_*=a/r_+,\;
 \Lambda^m_l=\sqrt{1+\ell}\big[l(l+1)-2ma\omega\big].\label{kx}
\end{eqnarray}
After using the redefinition $R(f)=f^{\alpha}(1-f)^{\beta}F(f)$, and the constraints
 \begin{eqnarray}\label{constrain}
\alpha^2+K^2_*=0,\;\beta^2-\beta+\big[K^2_*
-\Lambda^m_l\big]=0,
\end{eqnarray}
one can find that the equation (\ref{r1}) is a hypergeometric equation
\begin{eqnarray}
f(1-f)\frac{d^2F(f)}{d f^2}+[c-(1+\tilde{a}+\tilde{b})f]\frac{d F(f)}{d
f}-\tilde{a}\tilde{b} F(f)=0,\label{near2}
\end{eqnarray}
with
\begin{eqnarray}
\tilde{a}=\alpha+\beta,\quad
\tilde{b}=\alpha+\beta,\quad c=1+2\alpha.
\end{eqnarray}
 The above two constrains (\ref{constrain}) show that the parameters $\alpha$
and $\beta$ are as
\begin{eqnarray}
&&\alpha_{\pm}=\pm iK_*,\\
&&\beta_{\pm}=\frac{1}{2}\bigg[1\pm\sqrt{1-4\big(K^2_*
-\Lambda^m_l\big)} \;\bigg].\label{bet}
\end{eqnarray}
Then the exact analytical solution of Eq. (\ref{near2}) is
 \begin{eqnarray}
R(f)=A_-f^{\alpha}(1-f)^{\beta}F(\tilde{a}, \tilde{b}, c; f)+A_+
f^{-\alpha}(1-f)^{\beta}F(\tilde{a}-c+1, \tilde{b}-c+1, 2-c; f),
\end{eqnarray}
where $A_+, A_-$ are arbitrary constants.

 Near the horizon,
$r\rightarrow r_+$ and $f\rightarrow0$, the solution is
 \begin{eqnarray}
R_{NH}(f)=A_-f^{\alpha_\mp}+A_+f^{\alpha_\pm}.
\end{eqnarray}
The boundary condition near
the horizon is that there exists no outgoing mode, so we should to make either $A_-=0$ or $A_+=0$,
corresponding to the choice for $\alpha_\pm$. Here now we pick
$\alpha=\alpha_-$ and $A_+=0$. The convergence of the hypergeometric function
$F(\tilde{a}, \tilde{b}, c; f)$ will determine the sign of $\beta$, i.e. here Re$(c-\tilde{a}-\tilde{b})>0$, then we pick $\beta=\beta_-$ \cite{ding2010p,ding2010j}.
 Therefore the asymptotic solution near the horizon $(r\sim r_+)$ is
\begin{eqnarray}\label{near}
R_{NH}(f)=A_-f^{\alpha}(1-f)^{\beta}F(\tilde{a}, b, c; f).
\end{eqnarray}

To the
intermediate zone, we now can extend smoothly the near horizon solution (\ref{near}) to there.  We consider the property of the hypergeometric function \cite{mb} and
alter its argument in the near horizon solution from $f$ to $1-f$
\begin{eqnarray}
R_{NH}(f)&=&A_-f^{\alpha}(1-f)^{\beta}\bigg[\frac{\Gamma(c)\Gamma(c-\tilde{a}-\tilde{b})}
{\Gamma(c-\tilde{a})\Gamma(c-\tilde{b})}
F(\tilde{a}, \tilde{b}, \tilde{a}+\tilde{b}-c+1; 1-f)\nonumber\\
&+&(1-f)^{c-\tilde{a}-\tilde{b}}\frac{\Gamma(c)\Gamma(\tilde{a}+\tilde{b}-c)}{\Gamma(\tilde{a})\Gamma(\tilde{b})}
F(c-\tilde{a}, c-\tilde{b}, c-\tilde{a}-\tilde{b}+1; 1-f)\bigg].\label{r2}
\end{eqnarray}
When $r\gg r_+$, the function $(1-f)$ is
\begin{eqnarray}
1-f=\frac{2M}{r},
\end{eqnarray}
and then the near horizon solution (\ref{r2}) can be reduced
to
\begin{eqnarray}
R_{NH}(r)\simeq C_1r^{-\beta}+C_2r^{\beta-1}\label{rn2},
\end{eqnarray}
with
\begin{eqnarray}
&&C_1=A_-(2M)^{\beta}
\frac{\Gamma(c)\Gamma(c-\tilde{a}-\tilde{b})}{\Gamma(c-\tilde{a})\Gamma(c-\tilde{b})},\label{rn3}\\
&&C_2=A_-(2M)^{1-\beta}\frac{\Gamma(c)\Gamma(\tilde{a}+\tilde{b}-c)}{\Gamma(\tilde{a})\Gamma(\tilde{b})}.\label{rn4}
\end{eqnarray}

In the far field region, we can expand the
wave equation (\ref{radial}) as a power series in $1/r$ and maintain
only the leading items
\begin{eqnarray}
\frac{d^2R_{FF}(r)}{dr^2}+\frac{2}{r}\frac{dR_{FF}(r)}{d
r}+(1+\ell)\bigg[\omega^2-\frac{l(l+1)}{r^2}\bigg]R_{FF}(r)=0.
\end{eqnarray}
It easy to see that it is the usual Bessel equation. Therefore the solution of the radial
master equation (\ref{radial}) in the far-field limit can be
represented as
\begin{eqnarray}
R_{FF}(r)=\frac{1}{\sqrt{r}}\bigg[B_1J_{\nu}(\sqrt{1+\ell}\omega r)+B_2Y_{\nu}
(\sqrt{1+\ell}\omega r)\bigg],\label{rf}
\end{eqnarray}
where $J_{\nu}(\omega\;r)$ and $Y_{\nu}(\omega\;r)$ are the first
and second kind Bessel functions, $\nu=\sqrt{(1+\ell)l(1+l)+1/4}$. $B_1$ and
$B_2$ are integration constants. Now we can make the
limit $r\rightarrow 0$ and extend the far-field
solution (\ref{rf}) to small radial coordinate. Then the Eq.  (\ref{rf}) becomes
\begin{eqnarray}
R_{FF}(r)\simeq\frac{B_1(\frac{\sqrt{1+\ell}\omega r}{2})^{\nu}}{\sqrt{r}\;\Gamma(\nu+1)}
-\frac{B_2\Gamma(\nu)}{\pi
\sqrt{r}\;(\frac{\sqrt{1+\ell}\omega r}{2})^{\nu}}.\label{rfn2}
\end{eqnarray}
When applying the condition that the low-energy and low-angular momentum limit $(\omega r_+)^2\ll1$ and
$(a/r_+)^2\ll1$, the both power coefficients in
Eq.(\ref{rn2}) are approximately as
\begin{eqnarray}
-\beta &\simeq &-\frac{1}{2}+\nu + {\cal O}(\omega^2,a^2,a\omega), \\
(\beta-1)&\simeq &-\frac{1}{2}-\nu+ {\cal O}(\omega^2,a^2,a\omega).
\end{eqnarray}

Till now, it is easy to see that the both extensions (\ref{rn2}) and (\ref{rfn2}) of the near horizon and the far field solutions
can be simplified to power-law representations with the same power
coefficients, $r^{-1/2+\nu}$ and $r^{-1/2-\nu}$. By comparing the corresponding coefficients between
Eqs. (\ref{rn2}) and (\ref{rfn2}), we can get two connections
between $C_1,\;C_2$ and $B_1,\;B_2$. Eliminating $A_-$, we can gain
the ratio between the coefficients $B_1,\; B_2$
\begin{eqnarray}
B\equiv\frac{B_1}{B_2}&=&-\frac{1}{\pi}\bigg[\frac{1}{\sqrt{1+\ell}\omega M}\bigg]^{2\nu}
\nu\Gamma^2(\nu)
\nonumber\\
&\times&\;\frac{
\Gamma(c-\tilde{a}-\tilde{b})\Gamma(\tilde{a})\Gamma(\tilde{b})}{\Gamma(\tilde{a}+\tilde{b}-c)\Gamma(c-\tilde{a})\Gamma(c-\tilde{b})}.
\label{BB}
\end{eqnarray}

In the asymptotic region $r\rightarrow \infty$, the far-field solution
 can be rewritten as
\begin{eqnarray}
R_{FF}(r)&\simeq &
\frac{B_1+iB_2}{\sqrt{2\pi\;\sqrt{1+\ell}\;\omega} r}e^{-i\sqrt{1+\ell}\;\omega r}+
\frac{B_1-iB_2}{\sqrt{2\pi\;\sqrt{1+\ell}\;\omega} r}e^{i\sqrt{1+\ell}\;\omega r}\\
&=& A^{(\infty)}_{in}\frac{e^{-i\sqrt{1+\ell}\;\omega r}}{r}
+A^{(\infty)}_{out}\frac{e^{i\sqrt{1+\ell}\;\omega r}}{r}.\label{rf6}
\end{eqnarray}
The absorption probability can be obtained via the formula
\begin{eqnarray}
|\mathcal{A}_{l m}|^2=1-\bigg|\frac{A^{(\infty)}_{out}}{A^{(\infty)}_{in}}\bigg|^2
=1-\bigg|\frac{B-i}{B+i}\bigg|^2=\frac{2i(B^*-B)}{BB^*+i
(B^*-B)+1}.\label{GFA}
\end{eqnarray}
Substituting the expression of $B$ (\ref{BB}) into Eq.(\ref{GFA}), we
can acquire some features of absorption probability for the bumblebee
field coupled with Ricci tensor in the slowly rotating black hole
spacetime in the low-energy limit.

In Fig. 2, we set the angular momentum
$a=0.1$, and plot the variable of the absorption probability of a scalar
particle  for the first($l=0$) and second($l=1$) partial waves
in the slowly rotating Einstein-bumblebee black hole spacetime. It easy to  see
that  the absorption probability $A_{l=0}$
rises with the LV coupling constant $\ell$, which is analog to that of non-minimal coupling theory\cite{ding2010j}. However, $A_{l=1}$
lowers with the LV coupling constant $\ell$.
\begin{figure}[ht]\label{fig1}
\begin{center}
\includegraphics[width=8.0cm]{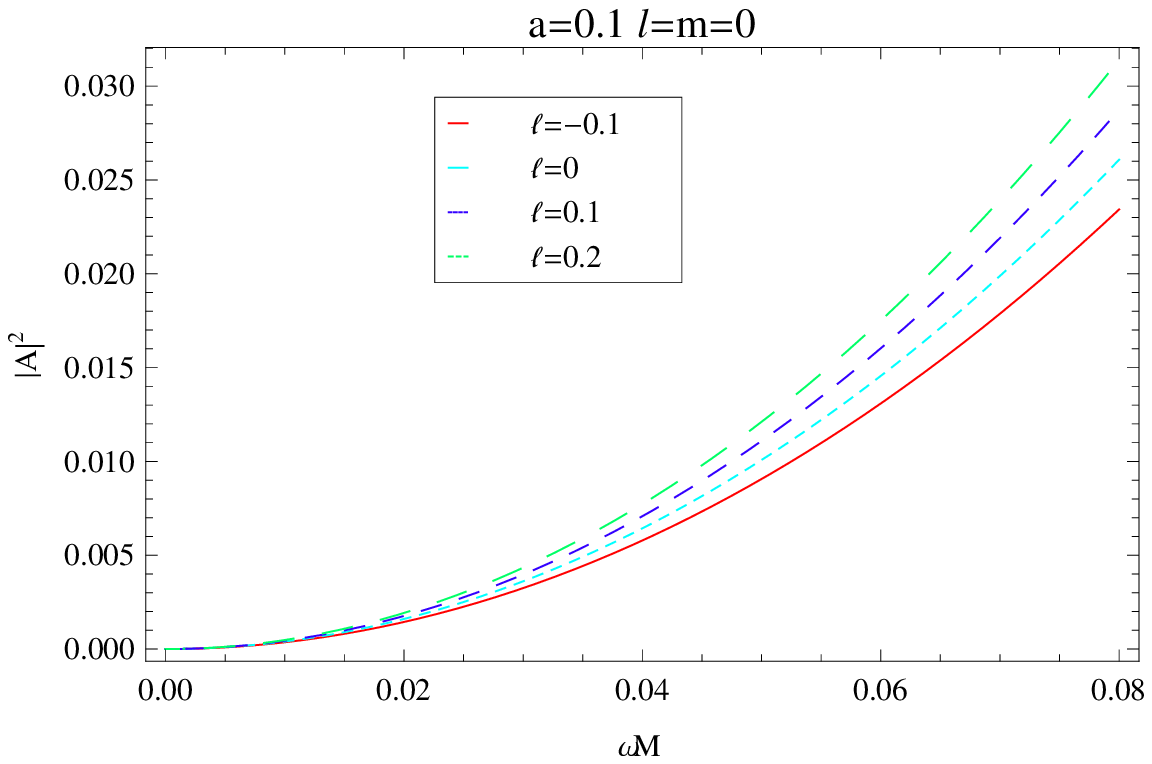}\;\;\;\;\includegraphics[width=8.0cm]{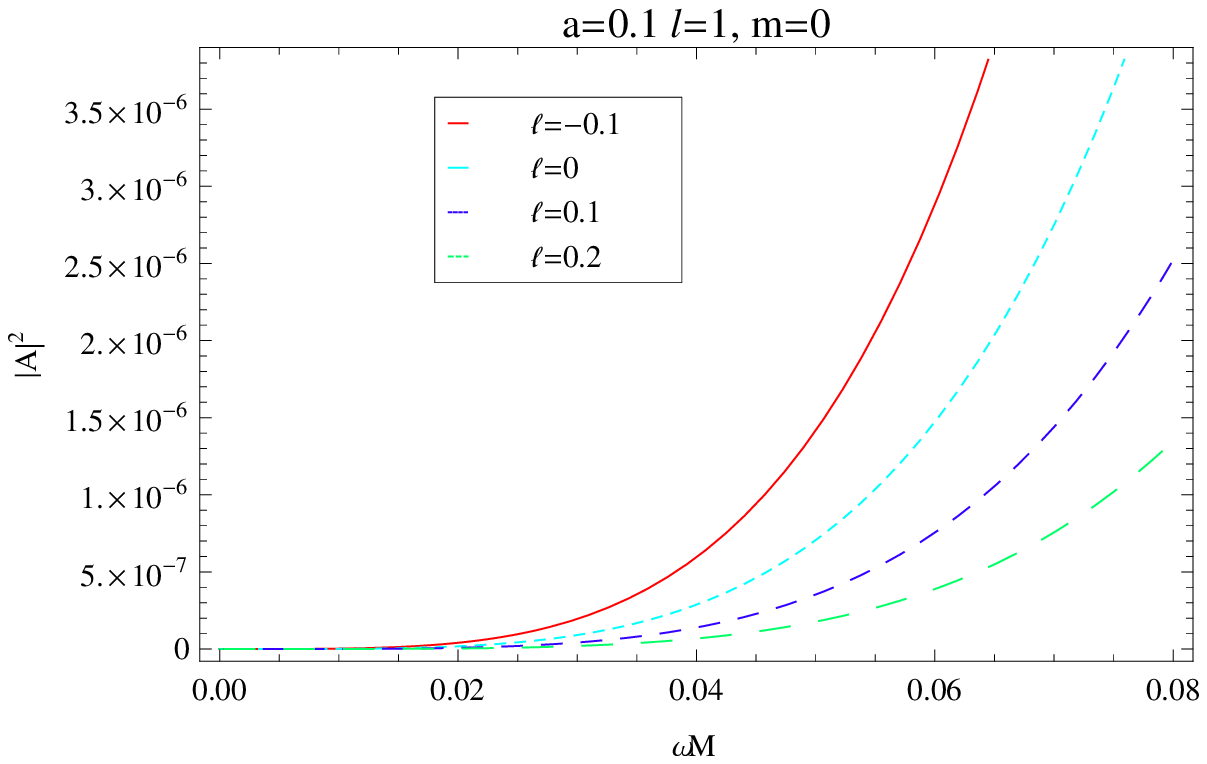}
\caption{Variety of the absorption probability $|A_{l m}|^2$ of a
 scalar field in the slowly rotating Einstein-bumblebee black hole for
 fixed  angular momentum $a=0.1$ and $m=0$.}
\end{center}
\end{figure}

 In the case of $l\geq 1$, there
has superradiation region when $m=1, 2, \cdots, l$, which is analogue to \cite{ding2010j}. In
Fig. 3, we showed the dependence of the absorption
probability on the angular index $l$ and $m$ with different $\ell$ and $a$. From the above two graphics in Fig. 3, for the super-radiation, the angular momentum $a$ improve the usual radiation ($m=-1$) and the super-radiation ($m=1$) for weak coupling.
Moreover, we see the attenuation
of $|A_{\ell m}|^2$ as the values of the angular index $l\geq 1$. This means
that the first partial wave $l=0$ leads over all others in the
absorption probability. It is similar to that of the scalar field
without any couplings.
\begin{figure}[ht]\label{fig2}
\begin{center}
\includegraphics[width=8.0cm]{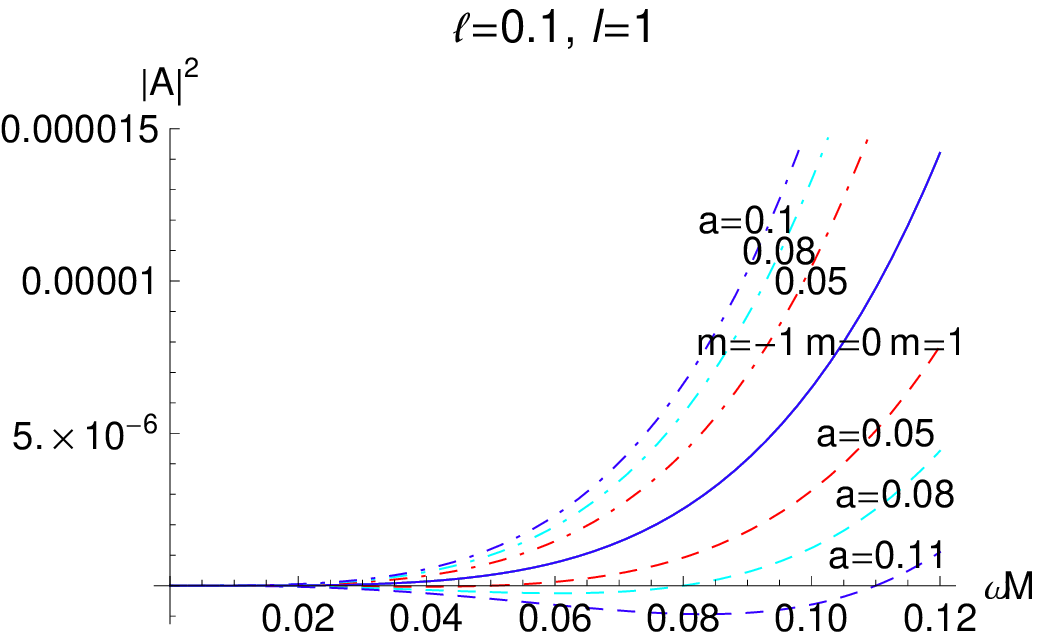}\;\;\;\;\includegraphics[width=8.0cm]{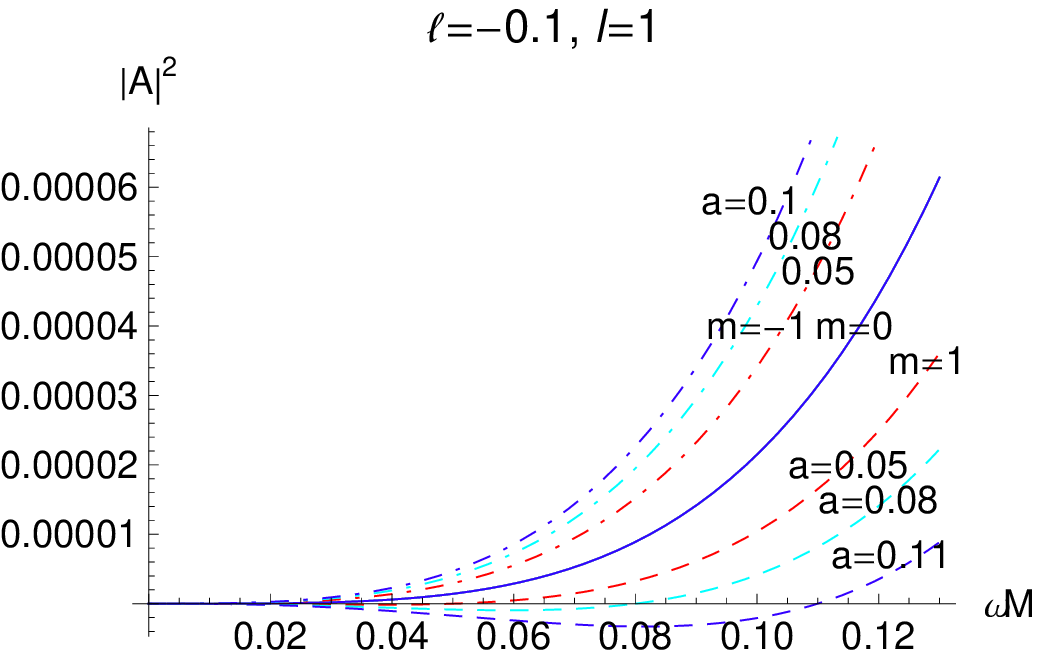}
\caption{The dependence of the absorption probability $|A_{l
m}|^2$ of a scalar field on the angular momentum $a$ in the
slowly rotating Einstein-bumblebee black hole for fixed $\ell=0.1,-0.1;l=1$ and
$m=1,0,-1$. }
 \end{center}
\end{figure}
\section{Summary}

In this paper, we have studied the slowly rotating, asymptotically flat black hole solutions of Einstein-bumblebee theory in the both cases of the bumblebee field $b_\mu=(0,b(r),0,0)$ and $b_\mu=(0,b(r),\mathfrak{b}(\theta),0)$.  In the case of radial Lorentz symmetry breaking, we have obtained an exact slowly rotating black hole solution to the $rr$ and $t\phi$ components of gravitational field equations.
When angular momentum $a\rightarrow0$, it can reduce to Schwarzschild like solution \cite{casana}; when LV constant $\ell\rightarrow0$, it can give the slowly rotating Kerr black hole solution. We then give the positions of horizon.

With this solution, we then check the other gravitational equations and the bumblebee motion equations. In the case of the bumblebee field $b_\mu=(0,b(r),0,0)$, all these equations can be fulfilled for arbitrary coupling constant $\ell$. In the cases of the bumblebee field $b_\mu=(0,b(r),\mathfrak{b}(\theta),0)$, all these equations can be fulfilled if and only if the coupling constant $\ell$ is as small as or smaller than the slowly rotating angular momentum $a$. If the LV coupling constant $\ell$ isn't small, the slowly rotating solution can't exist! Till now, there seems to be no full rotating black hole solution, so one can't use Newman-Janis algorithm to obtain a rotating solution, because it may not satisfy the whole set of field equations. It is similar to the Einstein-aether theory, where there can only exist a slowly rotating black hole solution.
It is an open issue that whether there exist a full rotating black hole solution for the case B when the coupling constant is very small.

With this obtained black hole solution, we can find some LV effects by the future astronomical events. So we study the black hole greybody factor for some observational effects of LV constant $\ell$. It shows that the deviation effects of LV from GR (slowly rotating Kerr black hole): when angular index $l=0$, the effective potential $V_{eff}$ decreases with the LV coupling constant $\ell$.  These decreases are similar to those of the Einstein-aether black hole \cite{tao} and non-minimal derivative coupling theory\cite{ding2010j}, which are also LV black hole. And the effect of the LV parameter on the greybody factor is that it enhances the absorption probability, which is similar to those of non-minimal derivative coupling theory\cite{ding2010j}. These difference could be detected by the new generation of gravitational antennas.

\begin{acknowledgments} The authors would thank Prof. Songbai Chen for useful suggestion. This work was supported by the Scientific Research Fund of the Hunan Provincial Education Department under No. 19A257, the National Natural Science Foundation (NNSFC)
of China (grant No. 11247013), Hunan Provincial Natural Science Foundation of China (grant No. 2015JJ2085).
\end{acknowledgments}
\appendix\section{Some quantities}
In this appendix, we showed the nonezero components of Ricci tensor for the metric (\ref{metric}). They are as following
\begin{eqnarray}
&&\mathcal{R}_{tt}=\frac{U}{1+\ell}\left[\frac{U''}{2}+\frac{r\sin^2\theta UU'}{\Sigma}\right]+\frac{a^2}{2\Sigma}\left[\frac{F^2U}{r^2}H'^2
+\frac{H^2U^2}{1+\ell}U'^2+\frac{F^2H^2U}{(1+\ell)r}U'-\frac{FH^2U}{1+\ell}F'U'\right]
,\\
&&\mathcal{R}_{t\phi}=-\frac{a}{2}\left[\frac{HUF''}{1+\ell}+\frac{FH''}{r^2}
+\frac{rFHU\sin^2\theta}{(1+\ell)\Sigma}U'-\frac{FU\cos\theta\sin\theta}{\Sigma}H'
\right]-\frac{a^3F^2H^3}{2\Sigma(1+\ell)}\left(\frac{1}{2}F'U'
+\frac{U}{r}F'\right),\\
&&\mathcal{R}_{rr}=-\frac{r^2\sin^2\theta}{\Sigma}\left(\frac{U''}{2}
+\frac{r\sin^2\theta}{\Sigma}UU'\right)-\frac{a^2F^2H^2}{\Sigma}F''+\frac{a^2H^2}{\Sigma^2}\Big[-\frac{1}{2}\bar\Sigma F'^2-\frac{F^2r^2\sin\theta}{4U}U'^2\nonumber\\
&&\qquad\quad
+\frac{F\bar\Sigma}{2U}F'U'+2rU\sin^2\theta FF'-\frac{F^2(5r^2U\sin^2\theta+a^2F^2H^2)}{2rU}U'-F^2U\sin^2\theta
\Big],\\
&&\mathcal{R}_{r\theta}=\frac{a^2FH}{\Sigma^2}
\Big[-\frac{1}{2}(a^2F^2H^2+3r^2U\sin^2\theta)F'H'+\frac{1}{2}Fr^2\sin\theta U'H'\nonumber\\
&&\qquad\quad+\frac{F}{r}(a^2F^2H^2+2r^2U\sin^2\theta)H'+\frac{1}{2}HUr^2\sin2\theta F'-FHr\sin\theta\cos\theta (\frac{1}{2}rU'+U)
\Big],\\
&&\mathcal{R}_{\theta\theta}=-\frac{r}{1+\ell}\left(U'
+\frac{r^3U^3\sin^4\theta}{\Sigma^2}\right)+\frac{r^4U^2\sin^4\theta}{\Sigma^2}-
\frac{a^2FH}{\Sigma}\Big[FH''-\frac{rFHU'}{2(1+\ell)}
+\frac{HrU}{1+\ell}F'\Big]\nonumber\\
&&\qquad\quad-\frac{a^2F^2}{\Sigma^2}\Big[\frac{1}{2}\bar\Sigma H'^2-HrU\sin2\theta H'+r^2H^2U\cos2\theta+\frac{r^2\sin^2\theta}{1+\ell}H^2U^2\Big],\\
&&\mathcal{R}_{\phi\phi}=-\frac{r^2U\sin^4\theta}{\Sigma}\left(
\frac{rU'+U}{1+\ell}-1\right)+
\frac{a^2}{\Sigma}\Big[-\frac{1}{2}F^2\sin^2\theta H'^2+\frac{1}{2}F^2H\sin2\theta H'\nonumber\\
&&\qquad\quad+
\frac{H^2\sin^2\theta}{1+\ell}\big(-\frac{1}{2}Ur^2F'^2
+\frac{1}{2}F^2rU'+FUrF'-2F^2U\big)-F^2H^2\cos2\theta\Big],
\end{eqnarray}
where $\Sigma$ and $\bar\Sigma$  are
\begin{eqnarray}
\Sigma=r^2U\sin^2\theta+a^2F^2H^2,\quad \bar\Sigma=r^2U\sin^2\theta-a^2F^2H^2.\nonumber
\end{eqnarray}
The nonzero components of the quantity $\bar B_{\mu\nu}$ are
\begin{eqnarray}
&&\bar B_{tt}=\frac{\ell U}{1+\ell}\left(\frac{U''}{2}
+\frac{U'}{r}\right)+\mathcal{O}(a^2),\\
&&\bar B_{t\phi}=-\frac{a\ell HUF''}{2(1+\ell)}-
\frac{a\ell FHU'}{r(1+\ell)}+\mathcal{O}(a^3),\\
&&\bar B_{rr}=0,\\
&&\bar B_{r\theta}=0,\\
&&\bar B_{\theta\theta}=-\frac{\ell}{1+\ell}\left(rU'+U\right)
+\mathcal{O}(a^2),\\
&&\bar B_{\phi\phi}=-\frac{\ell \sin^2\theta }{1+\ell}\left(
rU'+U\right)+\mathcal{O}(a^2),
\end{eqnarray}
for the case A; and
\begin{eqnarray}
&&\bar B_{tt}=\frac{\ell U}{1+\ell}\left(\frac{U''}{2}
+\frac{U'}{r}\right)+\frac{a\ell \cos2\theta\sqrt{U}}{2\sqrt{1+\ell}r^2\sin\theta}+\mathcal{O}(a^2),\\
&&\bar B_{t\phi}=-\frac{a\ell HUF''}{2(1+\ell)}-
\frac{a\ell FHU'}{r(1+\ell)}+\mathcal{O}(a^2),\\
&&\bar B_{rr}=0+\mathcal{O}(a^2),\\
&&\bar B_{r\theta}=\frac{a\ell\cos\theta}{r^2\sqrt{(1+\ell)U}}
(rU'+U)+\mathcal{O}(a^2),\\
&&\bar B_{\theta\theta}=-\frac{\ell}{1+\ell}\left(rU'+U\right)
+\frac{a\ell}{r\sin\theta\sqrt{(1+\ell)U}}(r\sin^2\theta U'-\cos2\theta U)+\mathcal{O}(a^2),\\
&&\bar B_{\phi\phi}=-\frac{\ell \sin^2\theta }{1+\ell}\left(
rU'+U\right)+
\frac{a\ell\sin\theta}{r\sqrt{(1+\ell)U}}\big(\frac{1}{2}rU'\cos^2\theta+\cos2\theta U\big)+\mathcal{O}(a^2),
\end{eqnarray}
for the case B.


\vspace*{0.2cm}
 \begin{thebibliography}{99}
 \baselineskip=0.6 cm



\bibitem{mattingly} D.~Mattingly,
Living Rev.\ Rel.\  {\bf 8}, 5 (2005).
\bibitem{camelia} G. Amelino-Camelia, Liv. Rev. Rel. {\bf 16}, 5 (2013).
\bibitem{dai} W.-M. Dai, Z.-K. Guo, R.-G. Cai and Y.-Z. Zhang, Eur. Phys. J. C {\bf77} 386 (2017).
\bibitem{colladay} D. Colladay and V.A. Kosteleck\'{y}, Phys. Rev. D {\bf55}, 6760 (1997); Phys. Rev. D {\bf58}, 116002 (1998); V. A. Kosteleck\'{y}, Phys. Rev. D {\bf69}, 105009 (2004).
\bibitem{coleman} S. R. Coleman and S. L. Glashow, Phys. Lett. B {\bf405}, 249 (1997); Phys. Rev. D {\bf59}, 116008 (1999); R. C. Myers and M. Pospelov, Phys. Rev. Lett. {\bf90},  211601 (2003).
\bibitem{rubtsov} G. Rubtsov, P. Satunin and, S. Sibiryakov, J. Cosmo. Astro. Phys. (JCAP) 05(2017)049.
\bibitem{kostelecky2004} V. A. Kosteleck\'{y}, Phys. Rev. D {\bf69}, 105009 (2004).
\bibitem{guiomar} G. Guiomar and J. P\'{a}ramos, Phys. Rev. D {\bf90}, 082002 (2014).
\bibitem{ding2015} C. Ding, A. Wang and X. Wang, Phys. Rev. D {\bf 92}, 084055 (2015).
\bibitem{ding2016} C. Ding, C. Liu, A. Wang and J. Jing, Phys. Rev. D {\bf 94}, 124034 (2016).
\bibitem{dickinson} M. H. Dickinson, F. O. Lehmann and S. P. Sane, Science {\bf284}, 1954 (1999).
\bibitem{kostelecky1989} V. A. Kosteleck\'{y} and S. Samuel, Phys. Rev. D {\bf40}, 1886 (1989).

\bibitem{casana} R. Casana and A. Cavalcante, Phys. Rev. D {\bf97}, 104001 (2018).
\bibitem{yang} Rong-Jia Yang, He Gao, Yao-Guang Zheng and Qin Wu, Commun. Theor. Phys. {\bf71}, 568 (2019).
\bibitem{ding2020} Chikun Ding, Changqing Liu, R. Casana and A. Cavalcate, Eur. Phys. C {\bf80}, 178 (2020).
\bibitem{bluhm}R. Bluhm, N. L. Gagne, R. Potting and A. Vrublevskis, Phys. Rev. D {\bf77}, 125007 (2008).
\bibitem{sch} K. Schwarzschild, Sitzungsber. Preuss. Akad. Wiss. (Math. Phys.) {\bf7}, 189 (1916).
\bibitem{kerr} R. P. Kerr, Phys. Rev. Lett. {\bf11}, 237 (1963).
\bibitem{newman} E. T. Newman and A. I. Janis, J. Math. Phys. {\bf6}, 915 (1965).
\bibitem{lammerzahl} C. L\"{a}mmerzahl, M. Maceda and A. Macias, Class. Quantum Grav. {\bf36}, 015001 (2019).
\bibitem{chen2020} Songbai Chen, Mingzhi Wang and Jiliang Jing, J. High Energy Phys. {\bf07}, 054 (2020).
\bibitem{barausse} E. Barausse, T. P. Sotiriou and I. Vega, Phys. Rev. D {\bf 93}, 044044 (2016).
\bibitem{ding2010p} C. Ding, S. Chen and J. Jing, Phys. Rev. D {\bf 82}, 024031 (2010).
\bibitem{ding2010j} C. Ding, C. Liu, J. Jing and S. Chen, J. High Energy Phys. {\bf11}, 146 (2010).
\bibitem{chen2015} S. Chen, J. Jing and H. Liao, Phys. Lett. B {\bf751},474 (2015).
 \bibitem{tao} T. Tao, Q. Wu, M. Jamil and K. Jusufi, Phys. Rev. D {\bf100}, 044055 (2019).
\bibitem{mb} M. Abramowitz and I. Stegun, \textit{Handbook of Mathematical
Functions}, Academic press, New York, U. S. A. (1996).
















\end{thebibliography}
\end{document}